\documentclass[12pt]{article}
\usepackage{a4}
\usepackage{latexsym} 
\usepackage{amssymb}  
\usepackage{amsmath} 
\usepackage{graphicx}
\usepackage{epsfig,wrapfig}
\usepackage[ usenames]{color}
\usepackage[rflt]{floatflt} 
\begin{document}

\newcommand{\talk}[3]
{\noindent{#1}\\ \mbox{}\ \ \ {\it #2} \dotfill {\pageref{#3}}\\[1.8mm]}
\newcommand{\stalk}[3]
{{#1} & {\it #2} & {\pageref{#3}}\\}
\newcommand{\snotalk}[3]
{{#1} & {\it #2} & {{#3}n.r.}\\}
\newcommand{\notalk}[3]
{\noindent{#1}\\ \mbox{}\ \ \ {\it #2} \hfill {{#3}n.r.}\\[1.8mm]}
\newcounter{zyxabstract}     
\newcounter{zyxrefers}        

\newcommand{\newabstract}
{\clearpage\stepcounter{zyxabstract}\setcounter{equation}{0}
\setcounter{footnote}{0}\setcounter{figure}{0}\setcounter{table}{0}}

\newcommand{\talkin}[2]{\newabstract\label{#2}\input{#1}}                

\newcommand{\rlabel}[1]{\label{zyx\arabic{zyxabstract}#1}}
\newcommand{\rref}[1]{\ref{zyx\arabic{zyxabstract}#1}}

\renewenvironment{thebibliography}[1] 
{\section*{References}\setcounter{zyxrefers}{0}
\begin{list}{ [\arabic{zyxrefers}]}{\usecounter{zyxrefers}}}
{\end{list}}
\newenvironment{thebibliographynotitle}[1] 
{\setcounter{zyxrefers}{0}
\begin{list}{ [\arabic{zyxrefers}]}
{\usecounter{zyxrefers}\setlength{\itemsep}{-2mm}}}
{\end{list}}

\renewcommand{\bibitem}[1]{\item\rlabel{y#1}}
\renewcommand{\cite}[1]{[\rref{y#1}]}      
\newcommand{\citetwo}[2]{[\rref{y#1},\rref{y#2}]}
\newcommand{\citethree}[3]{[\rref{y#1},\rref{y#2},\rref{y#3}]}
\newcommand{\citefour}[4]{[\rref{y#1},\rref{y#2},\rref{y#3},\rref{y#4}]}
\newcommand{\citefive}[5]
{[\rref{y#1},\rref{y#2},\rref{y#3},\rref{y#4},\rref{y#5}]}
\newcommand{\citesix}[6]
{[\rref{y#1},\rref{y#2},\rref{y#3},\rref{y#4},\rref{y#5},\rref{y#6}]}

\begin{center}
{\large\bf $\alpha_s$ with GAPP}\\[0.5cm]
Jens Erler \\[0.3cm]
Departamento de F\'isica Te\'orica, Instituto de F\'isica, \\
Universidad Nacional Aut\'onoma de M\'exico, 04510 M\'exico D.F., M\'exico
\end{center}

The FORTRAN package GAPP~\cite{Erler:1999ug} (Global Analysis of Particle Properties) computes so-called 
pseudo-observables and performs least-$\chi^2$ fits in the $\overline{\rm MS}$ scheme.
Fit parameters besides $\alpha_s$ and $M_H$ include the heavy quark masses which are
determined from QCD sum rule constraints thus affecting and being affected by $\alpha_s$.
When possible, analytical expressions (or expansions) are used to capture the full dependence on
$\alpha_s$ and the other fit parameters. 

$Z$-pole observables  from LEP~1 and SLC include the $Z$-width, $\Gamma_Z$, 
hadronic-to-leptonic partial  $Z$-width ratios, $R_\ell$, 
and the hadronic peak cross section, $\sigma_{\rm had}$.
These are most sensitive to $\alpha_s$ by far, 
but the weak angle enters and needs to be known independently.  
Thus, the extracted $\alpha_s$ depends on the set of other, purely electroweak (EW) measurements employed in the fits,
such as various asymmetries and experiments exploiting parity violation. 
The statistical and systematic experimental correlations of $\Gamma_Z$, $\sigma_{\rm had}$ 
and the $R_\ell$ are known, small and included. 
The parametric uncertainties (such as from $M_H$) are non-Gaussian but treated exactly. 
The theoretical errors in $\Gamma_Z$, $\sigma_{\rm had}$, and the $R_\ell$ are identical, 
and induce a negligibly small uncertainty in $\Delta\alpha_s(M_Z) = \pm 0.00009$, 
dominated ($\pm 0.00007$) by the axial-vector singlet contribution~\cite{Kniehl:1989bb} which is unknown at ${\cal O}(\alpha_s^4)$.
As in the case of $\tau$ decays, one may opt for either fixed-order perturbation theory (FOPT) or
contour-improved perturbation theory (CIPT)~\cite{Le Diberder:1992te}, and we take the 
difference\footnote{This difference has the opposite sign from $\tau$ decays indicating that their theory errors are uncorrelated.}
as the massless non-singlet uncertainty ($\pm 0.00005$).
The $W$-width also features a strong $\alpha_s$ dependence, but it is currently not competitive and usually interpreted
rather as a measurement of a combination of CKM matrix elements. 

The global EW fit excluding $\tau$ decays (the $Z$-pole alone)
yields $\alpha_s(M_Z) = 0.1203 \pm 0.0027$ ($0.1198 \pm 0.0028$).
These results are expected to be stronger affected by physics beyond the Standard Model than other $\alpha_s$ determinations
which is the primary reason to include another $\alpha_s$ constraint in the fits as a control.
If the new physics affects only the gauge boson propagators (oblique corrections) the resulting 
$\alpha_s(M_Z) = 0.1199^{+0.0027}_{-0.0030}$ hardly changes,
while allowing new physics corrections to the $Zb\bar{b}$-vertex gives the lower $\alpha_s(M_Z) = 0.1167 \pm 0.0038$.        

As the aforementioned $\alpha_s$ control we choose the $\tau$ lifetime, $\tau_\tau$, 
not least because of its transparent (even if controversial) theory uncertainty. 
Our master formula~\cite{Erler:2002bu} reads,
\begin{eqnarray}
\label{tau}
\tau^{\rm expt}
\equiv \tau [{\cal B}_{e,\mu}^{\rm expt}, \tau_{\rm direct}^{\rm expt}]
=  \hbar\, \frac{1 - {\cal B}_s^{\rm expt}}{\Gamma_e^{\rm theo} + \Gamma_\mu^{\rm theo} + \Gamma_{ud}^{\rm theo}} 
= 291.09 \pm 0.48~{\rm fs} \,,
\end{eqnarray}
where $\tau_{\rm direct}^{\rm expt} = 290.6~(1.0)$~fs is the directly measured $\tau$ lifetime~\cite{Nakamura:2010zzi}.
$\tau [{\cal B}_{e,\mu}^{\rm expt}] = 291.24~(0.55)$~fs is the combination of indirect determinations,
using $\tau [{\cal B}_{e, \mu}] = \hbar\, {\cal B}_{e,\mu}^{\rm expt}/\Gamma_{e,\mu}^{\rm theo}$
and the experimental branching ratios, ${\cal B}_e^{\rm expt} = 0.1785~(5)$ and ${\cal B}_\mu^{\rm expt} = 0.1736~(5)$,
together with their 13\% anti-correlation~\cite{Nakamura:2010zzi}.  
Decays into net strangeness, $S$, are plagued by the uncertainty in the  $\overline{\rm MS}$ strange mass, $\hat{m}_s (m_\tau)$,
and a poorly converging QCD series proportional to $\hat{m}_s^2$,
so that in Eq.~(\ref{tau}) we employ the measured $\Delta S = -1$ branching ratio, ${\cal B}_s^{\rm expt} = 0.0286~(7)$~\cite{Nakamura:2010zzi}.

The partial $\tau$-width into light quarks contains logarithmically enhanced EW corrections, 
$S(m_\tau, M_Z) = 1.01907 \pm 0.0003$~\cite{Erler:2002mv}, and reads (employing FOPT as advocated in Ref.~\cite{Beneke:2008ad}),
\begin{eqnarray}
\label{Gammaud}
\Gamma_{ud}^{\rm theo}
&=& \frac{G_F^2 m_\tau^5 |V_{ud}|^2}{64\pi^3} S(m_\tau, M_Z) \left(1 + \frac{3}{5} \frac{m_\tau^2}{M_W^2} \right)  \times \\ \nonumber
&  & \left( 1 + \frac{\alpha_s(m_\tau)}{\pi} + 5.202 \frac{\alpha_s^2}{\pi^2} + 26.37 \frac{\alpha_s^3}{\pi^3} + 127.1 \frac{\alpha_s^4}{\pi^4} 
- 1.393 \frac{\alpha(m_\tau)}{\pi} + \delta_q \right) \,,
\end{eqnarray}
where $\delta_q$ collects quark condensate, $\delta_{\rm NP}$~\cite{Maltman:2008nf}, as well as heavy and light quark mass effects.
The dominant experimental and theoretical errors are given in the following tables, respectively:
\begin{center}
\begin{tabular}{|c||c|r|}\hline
source                                       & uncertainty            & $\Delta\alpha_s(M_Z)$  \\ \hline\hline
$\Delta\tau^{\rm expt}$           & $\pm 0.48$~fs       & $\mp 0.00039$ \\ \hline 
$\Delta{\cal B}_s^{\rm expt}$ & $\pm 0.0007$       & $\mp 0.00017$ \\ \hline 
$\Delta V_{ud}$                       & $\pm 0.00022$     & $\mp 0.00007$ \\ \hline 
$\Delta m_\tau$                       & $\pm 0.17$~MeV & $\mp 0.00002$ \\ \hline \hline
total                                           &                                 & 0.00043             \\ \hline 
\end{tabular}
\hspace{42pt}
\begin{tabular}{|c||c|c|c|}\hline
source                                       & uncertainty                  &based on                      & $\Delta\alpha_s(M_Z)$  \\ \hline\hline
PQCD                                       & $\mp 0.0119$             & $\alpha_s^4$-term     & $^{+0.00167}_{-0.00137}$ \\ \hline 
RGE                                          & $\beta_4 = \mp 579$ & \cite{Erler:1999ug}     & $^{+0.00038}_{-0.00034}$ \\ \hline 
$\delta_{\rm NP}$                          & $\pm 0.0038$             &\cite{Maltman:2008nf} & $\mp 0.00048$ \\ \hline 
OPE \hspace{-32pt} ------  \hspace{32pt}& $\pm 0.0008$ & \cite{'tHooft:1976fv} \& \cite{Davier:2005xq} & $\mp 0.00012$ \\ \hline \hline
total                                           &                                      &                                        & $^{+0.00178}_{-0.00150}$ \\ \hline 
\end{tabular}
\end{center}
The perturbative QCD (PQCD) error dominates and is estimated as the $\alpha_s^4$-term in Eq.~(\ref{Gammaud}).
It is re-calculated in each call in the fits to access its $\alpha_s$-dependence and features asymmetric.
It basically covers the range from the higher values favored by CITP down to the lower ones one obtains from 
assuming that the roughly geometric form of FOPT continues.
Note that if CIPT is used, the error from the renormalization group evolution (RGE) parametrized 
by the unknown 5-loop $\beta$-function coefficient, $\beta_4$, and part of the PQCD error are correlated.
Effects breaking the operator product expansion, OPE \hspace{-32pt} ------, 
are estimated by assuming the instanton motivated functional form~\cite{'tHooft:1976fv}, 
$A\, \alpha_s^{-6} \exp[- 2\pi/\alpha_s(s_0)]$,
and adjusting $A$ to the difference between the OPE and data curves in Fig.~22 of Ref.~\cite{Davier:2005xq}.
Our result is $\alpha_s[\tau_\tau] = 0.1174^{+0.0018}_{-0.0015}$.

%


\begin{thebibliography}{99}

\bibitem{Erler:1999ug}
  J.~Erler,
  arXiv:hep-ph/0005084.\\[-5mm]
  
\bibitem{Kniehl:1989bb}
  B.~A.~Kniehl and J.~H.~K\"uhn,
  Phys.\ Lett.\  B224 (1989) 229.\\[-5mm]

  S.~A.~Larin, T.~van Ritbergen and J.~A.~M.~Vermaseren,
  Nucl.\ Phys.\  B438 (1995) 278.\\[-5mm]
 
\bibitem{Le Diberder:1992te}
  F.~Le Diberder and A.~Pich,
  Phys.\ Lett.\  B286 (1992) 147.\\[-5mm]
  
\bibitem{Erler:2002bu}
  J.~Erler and M.~Luo,
  Phys.\ Lett.\  B {\bf 558} (2003) 125.\\[-5mm]
  
\bibitem{Nakamura:2010zzi}
  Particle Data Group: K.~Nakamura {\it et al.},
  J.\ Phys.\ G37 (2010) 075021.\\[-5mm]

\bibitem{Erler:2002mv}
  J.~Erler,
  Rev.\ Mex.\ Fis.\  50 (2004) 200.\\[-5mm]
  
\bibitem{Beneke:2008ad}
  M.~Beneke and M.~Jamin,
  JHEP 0809 (2008) 044.\\[-5mm]

\bibitem{Maltman:2008nf}
  K.~Maltman and T.~Yavin,
  Phys.\ Rev.\  D78 (2008) 094020.\\[-5mm]

\bibitem{'tHooft:1976fv}
  G.~'t Hooft,
  Phys.\ Rev.\  D14 (1976) 3432.\\[-5mm]
  
\bibitem{Davier:2005xq}
  M.~Davier, A.~H\"ocker and Z.~Zhang,
  Rev.\ Mod.\ Phys.\ 78 (2006) 1043.\\[-5mm]
\end{thebibliography}
\end{document}